\documentclass[aps,prc,%
amsmath,amssymb,10pt,nofootinbib,showpacs,floatfix,showkeys,%
superscriptaddress%
]{revtex4-1}
\usepackage[utf8x]{inputenc}
\usepackage{graphicx}
\usepackage{multirow}

\begin{document}
\title{Lifetime measurements of $^{214}$Po and $^{212}$Po with the CTF liquid scintillator detector at LNGS.}
\author{G.~Bellini}
\affiliation{Dipartimento di Fisica, Universit\`{a} e INFN, Milano 20133, Italy}
\author{J.~Benziger}
\affiliation{Chemical Engineering Department, Princeton University, Princeton, NJ 08544, USA}
\author{D.~Bick}
\affiliation{Institut f\"ur Experimentalphysik, Universit\"at Hamburg, Germany}
\author{G.~Bonfini}
\affiliation{INFN Laboratori Nazionali del Gran Sasso, Assergi 67010, Italy}
\author{D.~Bravo}
\affiliation{Physics Department, Virginia Polytechnic Institute and State University, Blacksburg, VA 24061, USA}
\author{M.~Buizza Avanzini}
\affiliation{Dipartimento di Fisica, Universit\`{a} e INFN, Milano 20133, Italy}
\author{B.~Caccianiga}
\affiliation{Dipartimento di Fisica, Universit\`{a} e INFN, Milano 20133, Italy}
\author{L.~Cadonati}
\affiliation{Physics Department, University of Massachusetts, Amherst MA 01003, USA}
\author{F.~Calaprice}
\affiliation{Physics Department, Princeton University, Princeton, NJ 08544, USA}
\author{C.~Carraro}
\affiliation{Dipartimento di Fisica, Universit\`{a} e INFN, Genova 16146, Italy}
\author{P.~Cavalcante}
\affiliation{INFN Laboratori Nazionali del Gran Sasso, Assergi 67010, Italy}
\author{A.~Chavarria}
\affiliation{Physics Department, Princeton University, Princeton, NJ 08544, USA}
\author{A.~Chepurnov}
\affiliation{Institute of Nuclear Physics, Lomonosov Moscow State University, 119899, Moscow, Russia} 
\author{V.~Chubakov}
\affiliation{Dipartimento di Fisica, Universit\`{a} e INFN, I-44100 Ferrara, Italy}
\author{D.~D{\textquoteright}Angelo}
\affiliation{Dipartimento di Fisica, Universit\`{a} e INFN, Milano 20133, Italy}
\author{S.~Davini}
\affiliation{Dipartimento di Fisica, Universit\`{a} e INFN, Genova 16146, Italy}
\author{A.~Derbin}
\affiliation{St. Petersburg Nuclear Physics Institute, Gatchina 188350, Russia}
\author{A.~Etenko}
\affiliation{NRC Kurchatov Institute, Moscow 123182, Russia}
\author{K.~Fomenko}
\affiliation{INFN Laboratori Nazionali del Gran Sasso, Assergi 67010, Italy}
\affiliation{Joint Institute for Nuclear Research, Dubna 141980, Russia}
\author{D.~Franco}
\affiliation{APC, Univ. Paris Diderot, CNRS/IN2P3, CEA/Irfu, Obs de Paris, Sorbonne Paris Cité, France}
\author{C.~Galbiati}
\affiliation{Physics Department, Princeton University, Princeton, NJ 08544, USA}
\author{S.~Gazzana}
\affiliation{INFN Laboratori Nazionali del Gran Sasso, Assergi 67010, Italy}
\author{C.~Ghiano}
\affiliation{INFN Laboratori Nazionali del Gran Sasso, Assergi 67010, Italy}
\author{M.~Giammarchi}
\affiliation{Dipartimento di Fisica, Universit\`{a} e INFN, Milano 20133, Italy}
\author{M.~G\"{o}ger-Neff}
\affiliation{Physik Department, Technische Universit\"{a}t M\"{u}nchen, Garching 85747, Germany}
\author{A.~Goretti}
\affiliation{Physics Department, Princeton University, Princeton, NJ 08544, USA}
\author{L.~Grandi}
\affiliation{Physics Department, Princeton University, Princeton, NJ 08544, USA}
\author{E.~Guardincerri}
\affiliation{Dipartimento di Fisica, Universit\`{a} e INFN, Genova 16146, Italy}
\author{S.~Hardy}
\affiliation{Physics Department, Virginia Polytechnic Institute and State University, Blacksburg, VA 24061, USA}
\author{Aldo Ianni}
\affiliation{INFN Laboratori Nazionali del Gran Sasso, Assergi 67010, Italy}
\author{Andrea Ianni}
\affiliation{Physics Department, Princeton University, Princeton, NJ 08544, USA}
\author{V.~Kobychev}
\affiliation{Kiev Institute for Nuclear Research, Kiev 06380, Ukraine}
\author{D.~Korablev}
\affiliation{Joint Institute for Nuclear Research, Dubna 141980, Russia}
\author{G.~Korga}
 \affiliation{INFN Laboratori Nazionali del Gran Sasso, Assergi 67010, Italy}
\author{Y.~Koshio} 
 \affiliation{INFN Laboratori Nazionali del Gran Sasso, Assergi 67010, Italy}
\author{D.~Kryn}
\affiliation{APC, Univ. Paris Diderot, CNRS/IN2P3, CEA/Irfu, Obs de Paris, Sorbonne Paris Cité, France}
\author{M.~Laubenstein}
\affiliation{INFN Laboratori Nazionali del Gran Sasso, Assergi 67010, Italy}
\author{T.~Lewke}
\affiliation{Physik Department, Technische Universit\"{a}t M\"{u}nchen, Garching 85747, Germany}
\author{Marcello~Lissia}
\affiliation{Istituto Nazionale di Fisica Nucleare, Sezione di Cagliari, I-09042 Monserrato, Italy}
\author{E.~Litvinovich}
\affiliation{NRC Kurchatov Institute, Moscow 123182, Russia}
\author{B.~Loer}
\affiliation{Physics Department, Princeton University, Princeton, NJ 08544, USA}
\author{F.~Lombardi}
\affiliation{INFN Laboratori Nazionali del Gran Sasso, Assergi 67010, Italy}
\author{P.~Lombardi}
\affiliation{Dipartimento di Fisica, Universit\`{a} e INFN, Milano 20133, Italy}
\author{L.~Ludhova}
\affiliation{Dipartimento di Fisica, Universit\`{a} e INFN, Milano 20133, Italy}
\author{I.~Machulin}
\affiliation{NRC Kurchatov Institute, Moscow 123182, Russia}
\author{S.~Manecki}
\affiliation{Physics Department, Virginia Polytechnic Institute and State University, Blacksburg, VA 24061, USA}
\author{W.~Maneschg}
\affiliation{Max-Plank-Institut f\"{u}r Kernphysik, Heidelberg 69029, Germany}
\author{G.~Manuzio}
 \affiliation{Dipartimento di Fisica, Universit\`{a} e INFN, Genova 16146, Italy}
\author{Q.~Meindl}
\affiliation{Physik Department, Technische Universit\"{a}t M\"{u}nchen, Garching 85747, Germany}
\author{E.~Meroni}
\affiliation{Dipartimento di Fisica, Universit\`{a} e INFN, Milano 20133, Italy}
\author{L.~Miramonti}
\affiliation{Dipartimento di Fisica, Universit\`{a} e INFN, Milano 20133, Italy}
\author{M.~Misiaszek}
\affiliation{M. Smoluchowski Institute of Physics, Jagiellonian University, Krakow, 30059, Poland}
\author{D.~Montanari}
\affiliation{Fermi National Accelerator Laboratory, Batavia, IL 60510, USA}
\author{P.~Mosteiro}
\affiliation{Physics Department, Princeton University, Princeton, NJ 08544, USA}
\author{F.~Mantovani}
\affiliation{Dipartimento di Fisica, Universit\`{a} e INFN, I-44100 Ferrara, Italy}
\author{V.~Muratova}
\affiliation{St. Petersburg Nuclear Physics Institute, Gatchina 188350, Russia}
\author{S.~Nisi}
\affiliation{INFN Laboratori Nazionali del Gran Sasso, Assergi 67010, Italy}
\author{L.~Oberauer}
\affiliation{Physik Department, Technische Universit\"{a}t M\"{u}nchen, Garching 85747, Germany}
\author{M.~Obolensky}
\affiliation{APC, Univ. Paris Diderot, CNRS/IN2P3, CEA/Irfu, Obs de Paris, Sorbonne Paris Cité, France}
\author{F.~Ortica}
\affiliation{Dipartimento di Chimica, Universit\`{a} e INFN, Perugia 06123, Italy}
\author{K.~Otis}
\affiliation{Physics Department, University of Massachusetts, Amherst MA 01003, USA}
\author{M.~Pallavicini}
\affiliation{Dipartimento di Fisica, Universit\`{a} e INFN, Genova 16146, Italy}
\author{L.~Papp}
\affiliation{INFN Laboratori Nazionali del Gran Sasso, Assergi 67010, Italy}
\affiliation{Physics Department, Virginia Polytechnic Institute and State University, Blacksburg, VA 24061, USA}
\author{L.~Perasso}
\affiliation{Dipartimento di Fisica, Universit\`{a} e INFN, Milano 20133, Italy}
\author{S.~Perasso}
\affiliation{Dipartimento di Fisica, Universit\`{a} e INFN, Genova 16146, Italy}
\author{A.~Pocar}
\affiliation{Physics Department, University of Massachusetts, Amherst MA 01003, USA}
\author{G.~Ranucci}
\affiliation{Dipartimento di Fisica, Universit\`{a} e INFN, Milano 20133, Italy}
\author{A.~Razeto}
\affiliation{INFN Laboratori Nazionali del Gran Sasso, Assergi 67010, Italy}
\author{A.~Re}
\affiliation{Dipartimento di Fisica, Universit\`{a} e INFN, Milano 20133, Italy}
\author{A.~Romani}
\affiliation{Dipartimento di Chimica, Universit\`{a} e INFN, Perugia 06123, Italy}
\author{N.~Rossi}
\affiliation{INFN Laboratori Nazionali del Gran Sasso, Assergi 67010, Italy}
\author{A.~Sabelnikov}
\affiliation{NRC Kurchatov Institute, Moscow 123182, Russia}
\author{R.~Saldanha}
\affiliation{Physics Department, Princeton University, Princeton, NJ 08544, USA}
\author{C.~Salvo}
\affiliation{Dipartimento di Fisica, Universit\`{a} e INFN, Genova 16146, Italy}
\author{S.~Sch\"onert}
\affiliation{Physik Department, Technische Universit\"{a}t M\"{u}nchen, Garching 85747, Germany}
\affiliation{Max-Plank-Institut f\"{u}r Kernphysik, Heidelberg 69029, Germany}
\author{H.~Simgen}
\affiliation{Max-Plank-Institut f\"{u}r Kernphysik, Heidelberg 69029, Germany}
\author{M.~Skorokhvatov}
\affiliation{NRC Kurchatov Institute, Moscow 123182, Russia}
\author{O.~Smirnov}
\affiliation{Joint Institute for Nuclear Research, Dubna 141980, Russia}
\author{A.~Sotnikov}
\affiliation{Joint Institute for Nuclear Research, Dubna 141980, Russia}
\author{S.~Sukhotin}
\affiliation{NRC Kurchatov Institute, Moscow 123182, Russia}
\author{Y.~Suvorov}
\affiliation{INFN Laboratori Nazionali del Gran Sasso, Assergi 67010, Italy}
\affiliation{NRC Kurchatov Institute, Moscow 123182, Russia}
\author{R.~Tartaglia}
\affiliation{INFN Laboratori Nazionali del Gran Sasso, Assergi 67010, Italy}
\author{G.~Testera}
\affiliation{Dipartimento di Fisica, Universit\`{a} e INFN, Genova 16146, Italy}
\author{R.B.~Vogelaar}
\affiliation{Physics Department, Virginia Polytechnic Institute and State University, Blacksburg, VA 24061, USA}
\author{F.~von Feilitzsch}
\affiliation{Physik Department, Technische Universit\"{a}t M\"{u}nchen, Garching 85747, Germany}
\author{J.~Winter}
\affiliation{Physik Department, Technische Universit\"{a}t M\"{u}nchen, Garching 85747, Germany}
\author{M.~Wojcik}
\affiliation{M. Smoluchowski Institute of Physics, Jagiellonian University, Krakow, 30059, Poland}
\author{A.~Wright}
\affiliation{Physics Department, Princeton University, Princeton, NJ 08544, USA}
\author{M.~Wurm}
\affiliation{Physik Department, Technische Universit\"{a}t M\"{u}nchen, Garching 85747, Germany}
\author{G.~Xhixha}
\affiliation{Dipartimento di Fisica, Universit\`{a} e INFN, I-44100 Ferrara, Italy}
\author{J.~Xu}
\affiliation{Physics Department, Princeton University, Princeton, NJ 08544, USA}
\author{O.~Zaimidoroga}
\affiliation{Joint Institute for Nuclear Research, Dubna 141980, Russia}
\author{S.~Zavatarelli}
\affiliation{Dipartimento di Fisica, Universit\`{a} e INFN, Genova 16146, Italy}
\author{G.~Zuzel}
\affiliation{Max-Plank-Institut f\"{u}r Kernphysik, Heidelberg 69029, Germany}
\affiliation{M. Smoluchowski Institute of Physics, Jagiellonian University, Krakow, 30059, Poland}
\collaboration{Borexino Collaboration}
\date{4 Dicembre 2012 / Revised: 23 May 2013}

\begin{abstract}
We have studied the $\alpha$ decays of $^{214}$Po into $^{210}$Pb
and of $^{212}$Po into $^{208}$Pb tagged by the coincidence with the preceding
$\beta$ decays
from  $^{214}$Bi and $^{212}$Bi, respectively. The $^{222}$Rn, $^{232}$Th, and $^{220}$Rn 
sources used were sealed inside quartz vials and inserted in the Counting Test Facility
at the underground Gran Sasso National Laboratory in Italy. We find  that the mean lifetime of  $^{214}$Po is
$ (236.00 \pm 0.42 \mathrm{ (stat) }   \pm 0.15 \mathrm{ (syst) }  )\, \mu\mathrm{s}$ and
that of $^{212}$Po is
$ (425.1 \pm 0.9 \mathrm{ (stat) }   \pm 1.2 \mathrm{ (syst) }  )$~ns.
Our results,  obtained from data with signal--to--background ratio larger than 1000,
reduce the overall uncertainties and are compatible with previous measurements.
\end{abstract}

\maketitle

\section{Introduction}
Both $^{214}$Po and $^{212}$Po are polonium unstable isotopes, characterized by short mean-lives 
($\tau_{^{214}\rm{Po}} \approx 235 \,\mu\mathrm{s}$, $\tau_{^{212}\rm{Po}} \approx 430 \, \mathrm{ns}$) and by emission of alpha particles with 
energies $E_{^{214}\rm{Po}} \approx 7.833$~MeV and $E_{^{212}\rm{Po}} \approx 8.954$~MeV and practically 100\% branching ratios (BR). 
They belong to the $^{238}$U  and $^{232}$Th chains, respectively.  In this work, we report on the precise measurements of the $^{214}$Po 
and $^{212}$Po mean-lives, performed with the Counting Test Facility (CTF)~\cite{Alimonti:1998nt}, a $\approx 1$-ton liquid scintillator 
detector installed at the underground Gran Sasso National Laboratory (LNGS) in Italy.  In both measurements, 
we look for the fast coincidence
between the $\beta$-decay of the father isotope ($^{214}$Bi or
$^{212}$Bi) and the $^{214}$Po or $^{212}$Po $\alpha$-decay, respectively.
The $^{214}$Po lifetime 
measurement is characterized by unprecedented large statistics ($\approx 10^5$ events), and both the $^{212}$Po 
and $^{214}$Po measurements exploit a very long acquisition window ($\approx 7$  mean-lives).  Moreover, 
the high purity of the detector materials makes it possible to reach a signal--to--background ratio three orders of
magnitude better than the other existing measurements
listed in Tables~\ref{tab:bi214} and \ref{tab:bi212}.

\begin{table}[h]
 \caption[aaa1]{Existing measurements of the $^{214}$Po  half-life and compilations compared to the present work.
 \label{tab:bi214}}
\begin{tabular}{lc}
\hline \hline 
Reference & Half-life ($\mu$s)\\
\hline 
von Dardel (1950) \cite{vonDardel:1950} &   $163.7 \pm 0.2$ \\
Ballini (1953) \cite{Ballini:1953}  &  $158.0 \pm 2.0$ \\
Ogilvie (1960) \cite{Ogilvie:1960}  &  $159.5 \pm 3.0$  \\
Dobrowolski (1961) \cite{Dobrowolski:1961} &  $164.3 \pm 1.8$ \\
Erlik (1971) \cite{Erlik:1971} &  $165.5 \pm 3.0$ \\
Zhou (1993) \cite{Zhou:1993} &  $160.0 \pm 12.0$ \\
Alexeyev (2013) \cite{Alexeyev:2011rw} &  $163.5 \pm 0.8$\footnote{Average of 
the two values, 162.73 and 164.25, reported in ref. \cite{Alexeyev:2011rw}; since
systematic errors of these measurements are still under study \cite{Alexeyev:2011rw}
the uncertainty is estimated as $(164.25-162.73)/2$.} \\
\hline 
Nuclear Data Sheet (2009) \cite{Wu:2009} &  $164.3 \pm 2.0$\footnote{Average of \cite{vonDardel:1950},
\cite{Dobrowolski:1961} and \cite{Erlik:1971} with the original uncertainty of \cite{vonDardel:1950}  
increased to 2 $\mu$s; used in \emph{Table of Isotopes}.} \\
Table de Radionucl{\'e}ides (2007) \cite{Christe:2007} &  $162.3 \pm 1.2$\footnote{Average of all six values, 
with the original uncertainty of \cite{vonDardel:1950}  increased to 1.8 $\mu$s.} \\
\hline 
\textbf{This work (2012)}  &  $\mathbf{163.6 \pm 0.3}$ \\
\hline  \hline 
\end{tabular} 
\end{table}

\begin{table}[h]
\caption[aaa1]{Existing measurements of the $^{212}$Po half-life and compilations compared to the present work.}
  \label{tab:bi212}
\begin{tabular}{lc}
\hline \hline 
Reference & Half-life (ns) \\
\hline 
Bunyan (1949) \cite{Bunyan:1949}  &   $304 \pm 4$  \\
Flack (1962) \cite{Flack:1962} &  $305 \pm 25$  \\
Astner (1963) \cite{Astner:1963} &  $305 \pm 5$  \\
McBeth (1972) \cite{McBeth:1972} &  $304 \pm 8$\footnote{Curve (A) of Fig. 4 in Ref. \cite{McBeth:1972}.}\\
McBeth (1972) \cite{McBeth:1972} &  $300 \pm 8$\footnote{Curve (B) of Fig. 4 in Ref. \cite{McBeth:1972}.}\\
Sanyal  (1975) \cite{Sanyal:1975} &  $296 \pm 2$ \\
Bohn  (1981) \cite{Bohn:1981} &  $309 \pm 11$ \\
\hline 
Table of Isotopes (2005) \cite{Browne:2005} &  $299 \pm 2$\footnote{Average of  \cite{Bunyan:1949}, 
\cite{Astner:1963}, \cite{McBeth:1972} (curve (A) of Fig. 4), \cite{Sanyal:1975}, and \cite{Bohn:1981}.}
\\
Table de Radionucl{\'e}ides (2010) \cite{Nichols:2004} &  $300 \pm 2$\footnote{Average of \cite{Bunyan:1949}, 
 \cite{Flack:1962}, \cite{Astner:1963}, \cite{McBeth:1972} (both curves (A) and (B) of Fig. 4), \cite{Sanyal:1975},
 and \cite{Bohn:1981},  with the original uncertainty of  \cite{Sanyal:1975}  increased to 2.7 ns.}\\
\hline 
\textbf{This work (2012) first source}  &  $\mathbf{295.6 \pm 1.3}$ \\
\textbf{This work (2012) second source}  &  $\mathbf{294.0 \pm 1.1}$ \\
\textbf{This work (2012) combined}  &  $\mathbf{294.7 \pm 1.0}$ \\
\hline  \hline 
\end{tabular} 
\end{table}

\section{The experimental setup}
\label{TheExperimentalSetup}

\subsection{The measurement principle}
\label{Detection principle}

The $\alpha$ particles from $^{212}$Po and $^{214}$Po decays are tagged via the coincidence with $\beta$'s from 
$^{212}$Bi and $^{214}$Bi. The $^{212}$Bi is characterized by a $\beta$--decay with $\approx $~64\%~BR and $Q$--value 
of $\approx $~2.254~MeV, while $^{214}$Bi $\beta$--decay has almost 100\% BR and 
3.272~MeV $Q$--value. The times elapsed between the prompt $\beta$ and delayed $\alpha$ events are driven by the decay times of 
polonium and are therefore short (a few hundreds  microseconds for $^{214}$Po and 
a few hundreds  nanoseconds for $^{212}$Po). The CTF electronics are able to detect such time differences with high precision. 
The space--time coincidence condition provides a very clean event signature. 
In comparison with other measurements of this kind, the background level is further suppressed by the high radio-purity of the CTF detector~\cite{Alimonti:1998b}.

\subsection{The Counting Test Facility}
\label{The Counting Test Facility (CTF)}

A detailed  description of CTF can be found in~\cite{Alimonti:1998nt}, while here we only recall  those detector features
that are important for the particular measurement presented in this paper. The external cylindrical tank ($\approx 11$~m diameter, 
$\approx 10$~m height) was filled with about 1000 tons of water serving as a passive shield against external neutrons and $\gamma$'s. 
The core of the detector was represented by
 4.8~m$^{3}$ of liquid organic scintillator (LS)\footnote{Binary mixture of the aromatic liquid 1-2-4 trimethyl benzene 
(pseudocumene or PC) as solvent and the fluor admixture of 2,5--diphenyloxasol (PPO) with about 1.5 g/l concentration.} 
contained in a spherical inner vessel (IV) of $\approx 2$~m diameter. The IV was realized in $\approx 500$~$\mu$m thick nylon membrane with 
excellent optical clarity, which allowed the effective transmission of the scintillation light to the 100 photomultiplier tubes
 (PMTs) performing the optical read--out. These 8 inch ETL 9351 PMTs, which were anchored on a 7~m diameter support structure placed inside the water 
tank, had $\approx 26$\% quantum efficiency at 420~nm, a limited transit time spread ($\sigma \approx1$~ns), and a 
good pulse height resolution for single photoelectron pulses (Peak/Valley $\geq$ 2.5). Light concentrators (57~cm long and 50~cm diameter aperture) mounted on the PMTs enhanced the optical coverage to about 20\%. 

The fluorescence maximum is at 365~nm. 
The yield of emitted photons was $\approx 10^{4}$ per MeV of deposited energy, while the measured light yield corresponded on 
average to 380 photoelectrons (p.e.) collected by 100 PMT's per 1~MeV of energy deposit.
For each trigger (event), the charge and timing (1~ns precision) of all hit PMTs were recorded. Each electronics channel was paired with an auxiliary 
channel of equal characteristics, able to record a possible second event occurring within an 8~ms time window.  The parameters reconstructed for each 
event were: the total charge collected by the PMTs during a 500~ns time interval in order to infer the energy deposit, the time distribution of hit 
PMTs used for the pulse shape discrimination and  for the event position reconstruction (resolution of 10 -- 15~cm), and the time elapsed between 
sequential events, used for the precise time difference measurement between correlated events.

A special insertion system installed on the top of the CTF made it possible to place suitable radioactive sources inside 
the detector without contaminating the LS. The pipe of the insertion system had an internal diameter of 50~mm. The positioning of the sources 
was done by means of steerable aluminum rods. The $^{214}$Po and $^{212}$Po sources were placed at the detector center and its exact positions were 
later determined via the position reconstruction code which had a precision of a few centimeters.

\subsection{Preparation of the sources}
\label{subsec:PreparationOfTheSources}
The $^{214}$Po and $^{212}$Po sources were prepared by dissolving the suitable isotopes in the LS, which was contained in the vials with an external diameter of 50 mm, the maximum allowed by the insertion system. The vials were made of quartz, which is transparent to  UV light. The LS used in the source preparation was drawn directly from the CTF inner vessel in order to ensure the same optical 
properties in terms of the light yield and quenching effects. In order to minimize a possible contamination with Oxygen, which acts as a quencher with a consequent  light--yield reduction, the LS withdrawal was performed under a controlled high--purity Nitrogen atmosphere~\cite{Zuzel:2004}. 
These sources have been prepared with the purpose of studying  anti-neutrino spectra important in geoneutrino studies as discussed in ~\cite{Fiorentini:2009bv}.

\subsubsection{Preparation of the $^{214}$Po source}

The vial employed for the $^{214}$Po source was a 50~mm diameter sphere
with $\approx  1$~mm wall thickness. The source was realized by spiking the LS with $^{222}$Rn obtained from a $^{226}$Ra--based ``Radon generato'' and having
the $^{214}$Bi--$^{214}$Po sequence within its decay chain. The $^{222}$Rn
half-life of 3.824~days is long enough to allow the realization of an effective spiking procedure.  
To maintain the mandatory Oxygen--free conditions of the scintillator, we built a dedicated system 
allowing the Rn influx into the vial under a strictly controlled flow of high--purity Nitrogen.
The Rn solubility in the PC is high enough to allow an efficient capture of the radioisotope into the scintillator.

The required total source activity was about 10~Bq. Such a low counting rate matches the electronics maximum read-out
(the electronics dead-time is 110--120 ms, slightly dependent on energy) and also excludes the possibility of saturation and/or pile-up effects. Such a low--activity source was obtained by dissolving a few ml of 8000~Bq mother solution (Radon dissolved in PC) in the LS. 
The mother solution can have higher Oxygen contamination than the original LS. The concept of high--activity mother solution makes it possible to minimize the amount used in the source preparation. The data taking started immediately after inserting the 10~Bq source in the CTF and it was stopped when the source activity decayed to a few Bq.

\subsubsection{Preparation of the two $^{212}$Po sources}

The mere replication of the procedure adopted for $^{214}$Po was not possible, since
 $^{220}$Rn, the radon progenitor of the $^{212}$Bi, has a half--life of only 55.6~s. 
Two alternative methods, described below, were used instead in order to prepare two different $^{212}$Po sources.

{\it A) First $^{212}$Po source}

In this approach we used the fact that $^{212}$Po is a member of the $^{232}$Th decay chain. We employed natural thorium dissolved in Nitric Acid at 2\%. Since thorium is insoluble in PC, we had to use TriButyl Phosphate (TBP) to form stable hydrophilic complexes of thorium. These compounds are soluble in organic solvents and were mixed in the scintillator that was inserted into the vial. The concentration of thorium in TBP was measured by 
inductively-coupled plasma mass spectrometry
to be about 100~ppb corresponding roughly to 43~Bq/kg. The TBP concentration was kept below 5\% according to fluorimetric measurements  to minimize quenching effects.  In order to accumulate sufficient statistics, we increased the source volume; we built a cylindrical quartz vial 
20~cm high and with the maximum diameter allowed by the insertion system for a total volume of about 300~cm$^{3}$.

The $^{212}$Po source had a total activity of about half a Bq; it did not change significantly during the data taking, which lasted a 
short time compared  to the $^{232}$Th half--life of about 14 billion years.

{\it B) Second $^{212}$Po source}

As an alternative method avoiding the TBP dilution into the scintillator, a second source was prepared using a system able to transfer 
$^{220}$Rn gas directly into the spherical quartz vial. We again used natural thorium dissolved in Nitric Acid at 2\%. The core of the 
system was an extraction chamber, a 40~cm high stainless steel cylinder with 16~mm internal diameter, containing 20~cm$^3$ of aqueous solution 
with $\approx 1$~g/l of dissolved natural thorium with 80~Bq of total activity. High--purity Nitrogen was fluxed into the extraction chamber, in 
order to bubble the aqueous solution and to transfer the emanated $^{220}$Rn to the scintillator contained in the vial. 
The high purity Nitrogen loaded with $^{220}$Rn was flushed inside a 7~m long nylon tube with~2 mm internal diameter. Because of its short lifetime, 
only a fraction of the original $^{220}$Rn reached the vial. A nylon trap was inserted on top of the extraction chamber to avoid potential water 
contamination of the scintillator. The 2~mm nylon tube was contained inside a second nylon tube of 6~mm internal diameter, that acted both as 
Nitrogen exhaust and as a tether for the quartz vial. Since both tubes were flexible, a 3~kg stainless--steel cylindrical weight was positioned just 
above the vial in order to hold its position in the CTF center. The Nitrogen flux was set to 10~liters per hour with a corresponding gas velocity 
inside the 2~mm nylon of about 90~cm/s. The expected $^{212}$Bi activity in the vial was about 4~Bq. Given that the half-life of $^{212}$Pb is about 
10.6~h, the desired $^{212}$Po activity was built up after approximately 20-30 hours.

\section{Data analysis and results}

The total live-time of data taking for the $^{214}$Po measurement is 10.2~d. For the two $^{212}$Po sources, the live-times are 6.3~d and 15.5~d,
 respectively. The correspondent overall statistics are $\approx 3.9\times 10^5$ $^{214}$Po  and  $\approx (1.1+1.7)\times 10^5$ $^{212}$Po 
candidates for the first and second source, respectively. 
The two $^{212}$Po candidate samples from the two sources  are analyzed  independently.

\begin{figure}
 \centering
 \includegraphics[width=0.8\linewidth]{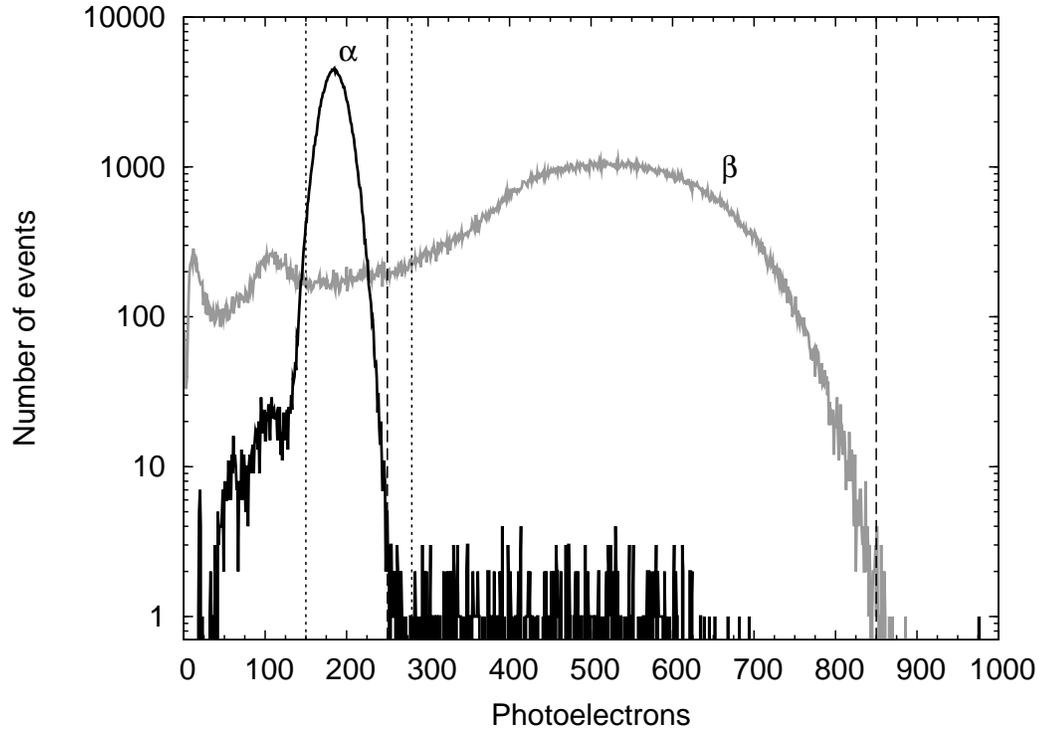}
 \caption{Energy spectrum for the prompt (candidate $\beta$ decay of $^{214}$Bi, gray histogram) and delayed event 
(candidate $\alpha$ decay of $^{214}$Po, black histogram) of TCE. Vertical dashed (dotted) lines delimit energy cuts for the $\beta$ ($\alpha$) candidates  used in the analysis. }
 \label{fig:Bi214spectrum}
\end{figure}

\begin{figure}
 \centering
 \includegraphics[width=0.8\linewidth]{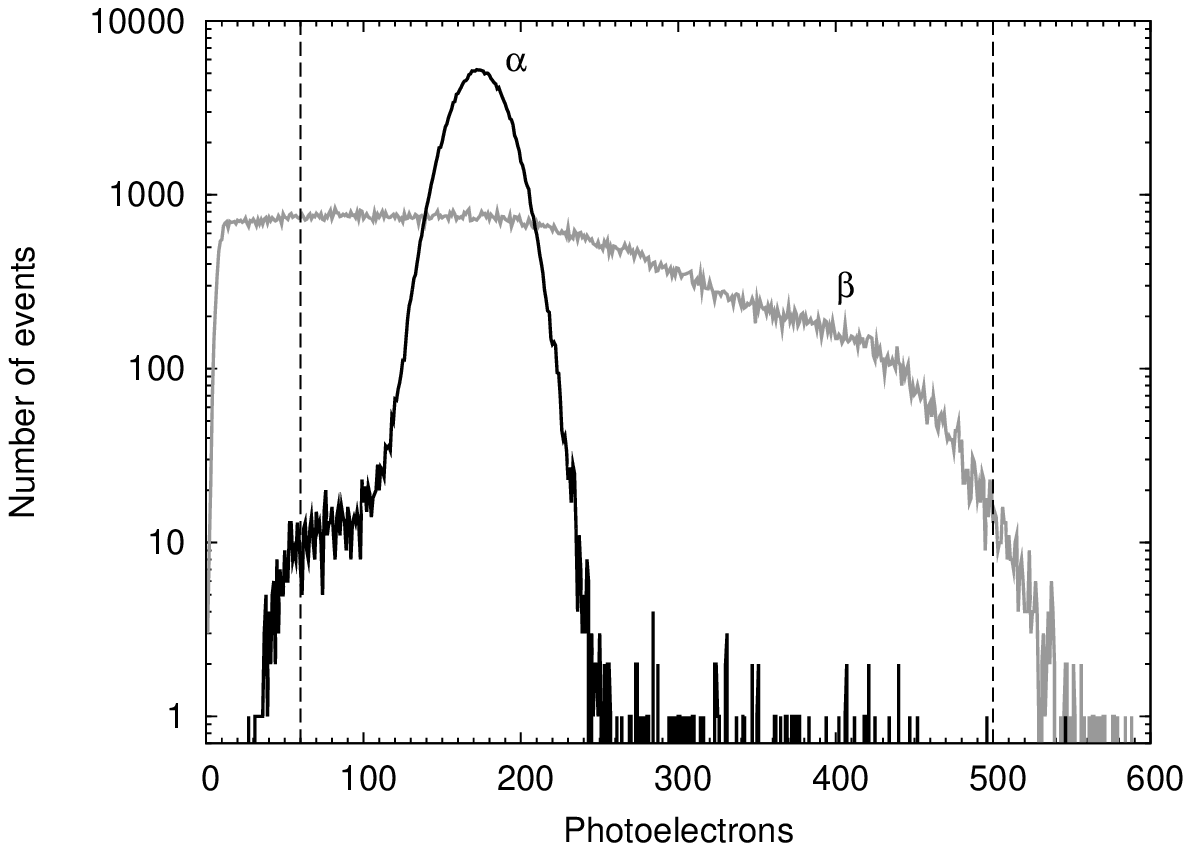}
 \caption{Energy spectrum for the prompt (candidate $\beta$ decay of $^{212}$Bi, gray histogram) and delayed event (candidate $\alpha$ decay of $^{212}$Po, black histogram) of TCE from the first source. 
Vertical dashed lines delimit the energy cuts  used in the analysis for the $\beta$ candidate; no cuts were used for the second event. }
 \label{fig:Bi212spec1}
\end{figure}

\begin{figure}
 \centering
 \includegraphics[width=0.8\linewidth]{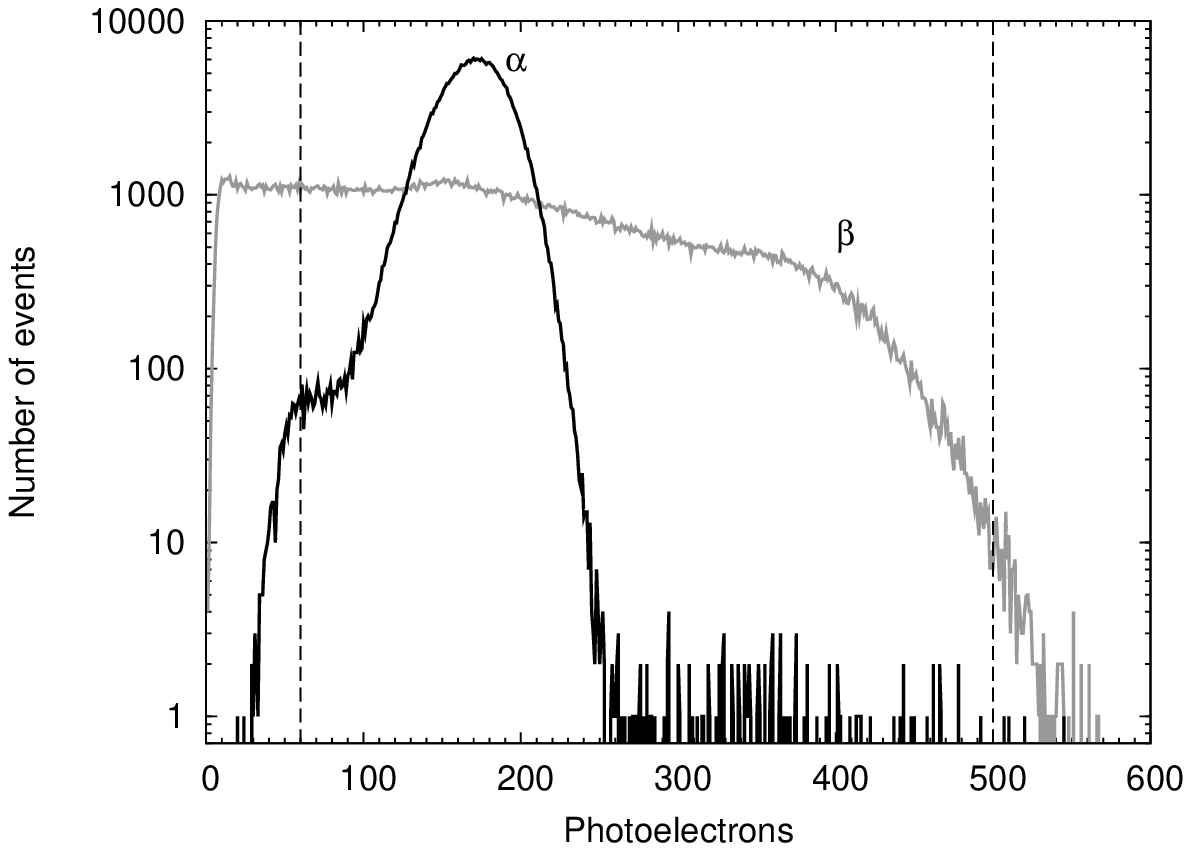}
 \caption{Energy spectrum for the prompt (candidate $\beta$ decay of $^{212}$Bi, gray histogram) 
and delayed event (candidate $\alpha$ decay of $^{212}$Po, black histogram) of TCE from the second $^{212}$Po source. 
Vertical dashed lines delimit the energy cuts  used in the analysis for the $\beta$ candidate; no cuts were used for the second event.}
 \label{fig:Bi212spec2}
\end{figure}

The energy response of the detector is calibrated run-by-run, using the light yield obtained
by fitting the $^{14}$C energy spectrum: on average 3.8
p.e. per PMT are detected for a 1~MeV recoiling electron at a random position within the detector
volume~\cite{Bellini:2008zza}.
 In these measurements, the  yields are 240 p.e./MeV  and 220 p.e./MeV with 70 PMTs for the $^{214}$Po and  $^{212}$Po analyses, respectively. 
The reduced yields are due to impurities, such as Oxygen in the $^{214}$Po vial and thorium salts in the $^{212}$Po vial,
which act as light quenchers.
The $^{214}$Bi -- $^{214}$Po and $^{212}$Bi -- $^{212}$Po (source 1 and 2) energy spectra are shown in 
Figures~\ref{fig:Bi214spectrum},  \ref{fig:Bi212spec1}, and  \ref{fig:Bi212spec2}. 
These figures also show that the measured light yield from alpha particles (polonium decays) are quenched by 
a factor $\approx 10$--15 with respect to electron equivalents. 

A time threshold of 400~ns for the $^{212}\mathrm{Bi}\to {}^{212}\mathrm{Po}$ decay sample was imposed to avoid the 
scintillation tail of the first event. A threshold of 5~$\mu$s was applied for  the  $^{214}\mathrm{Bi}\to {}^{214}\mathrm{Po}$ 
decay sample in  order to avoid  background 
from $^{212}\mathrm{Bi}\to {}^{212}\mathrm{Po}$ decays with a lifetime of about 0.4~$\mu$s, naturally present in the scintillator. 

The data analysis relies on a triple approach in order to cross check the results and to minimize the errors. Results from these approaches
are summarized in Table~\ref{tab:res} and described below.

\begingroup
\begin{table}[h]
\caption[aaa4]{Summary of mean-life results. Final results come from the mean value approach with energy cuts and also report
systematic errors; the $^{212}$Po results include a $-0.4$~ns correction for systematics as
discussed in the text. \label{tab:res}}
\begin{tabular}{l@{\hskip 0.8cm}c@{\hskip 0.8cm}c@{\hskip 0.8cm}c@{\hskip 0.8cm}c@{\hskip 0.8cm}c}
\hline \hline 
       &$^{214}$Po    [$\mu$s]        & \multicolumn{2}{c} {$^{212}$Po  [ns] }  &   \\
Method &          & source 1 & source 2&   \\
\hline
Analytical fit                           & 236.26 $\pm$ 0.47 &                  &                 \\
$\chi^2$ with penalty                    &                   & 426.8 $\pm$ 1.3  & 424.2 $\pm$ 1.1 \\
$\chi^2$ with penalty and energy cuts    &  236.00 $\pm$ 0.43                   & 426.5 $\pm$ 1.4        &  424.1 $\pm$ 1.2                \\
Mean value                               & 236.85 $\pm$ 0.42 & 426.8 $\pm$ 1.5  & 424.2 $\pm$ 1.1 \\
Mean value  with energy cuts             & 236.00 $\pm$ 0.42                    & 426.5 $\pm$ 1.4         & 424.2 $\pm$ 1.1                 \\
\hline 
\multirow{2}{*}{Final result} & \multirow{2}{*}{236.00 $\pm$ 0.42 $\pm$  0.15} & 426.5 $\pm$ 1.4  $\pm$  1.2 & 424.2  $\pm$  1.1  $\pm$  1.2 \\
\cline{3-4}
                                         &                        & \multicolumn{2}{c} {425.1 $\pm$ 0.9 $\pm$ 1.2}  \\
 
\hline  \hline 
\end{tabular} 
\end{table}
\endgroup

\subsection{$\chi^2$ with penalty factor}
\label{subsec:penalty}
In order to suppress the background contamination, polonium candidates are selected by looking for pairs of time correlated events (TCE), in time windows 
of 1.7~ms for the $^{214}$Po  and 3~$\mu$s for the $^{212}$Po analysis. 
Potential contaminations due to random coincidences are further reduced by applying energy cuts: [250 -- 850] and [150 -- 280] p.e. for the  
$^{214}$Bi and $^{214}$Po (Fig.~\ref{fig:Bi214spectrum}), and [60 -- 500] p.e. for the $^{212}$Bi and no cut for $^{212}$Po 
(Fig.~\ref{fig:Bi212spec1} and  \ref{fig:Bi212spec2}).
These energy cuts reduce the background by about two orders of magnitude in the $^{214}$Po analysis, and by a factor $\approx 2$ for $^{212}$Po.
 The distributions of the measured time differences between the two events, with and without the energy cuts, are shown in 
Figures~\ref{fig:decayIstogram214} and \ref{fig:decayIstogram212}, respectively.  

The mean lifetime is then evaluated by minimizing:

\begin{equation}
\label{eq3}
\chi^2(n_0,\lambda,b_0) = \chi^2_0(n_0,\lambda,b_0) + \left(\frac{b_0-\bar{b}}{\Delta\bar{b}}\right)^2
\end{equation}
where n$_0$ is the number of polonium decays, $\lambda$ the decay constant, and $b_0$ the   background due to accidental coincidences.
The $\chi^2_0(n_0,\lambda,b_0)$ is estimated by using the exponential model:

\begin{equation}
b_0 +  n_0 \exp(-\lambda t) .
\label{eq:expPlusBack}
\end{equation}
The expected value of the background $\bar{b}$ with  error $\Delta\bar{b}$ is estimated by 
measuring it independently in the off--time 
windows ([3.0 -- 7.6]~ms for the $^{214}$Po case, and [8.0 -- 48.0]~$\mu$s for the $^{212}$Po one, scaled to the length of the time windows 
used for the measurements ([0.005 -- 1.705]~ms and [0.4--3.4]~$\mu$s, respectively). 
Since the electronics acquire only the first event following 
the bismuth candidate, the  background related component is Poissonian and has an exponential behavior: 
\begin{equation}
\label{eq:background}
   b_0 e^{-b_1 t},
\end{equation}
where  $b_1$ is the background rate. However, all our background data samples do not show 
deviations from the constant accidental background ($b_1=0$ within the errors in the 
selected time windows), as expected given the low background rate.

Total-number-of-events/estimated-background-events in the selected time windows
are about $3.1\times 10^5 / 17$ with energy cuts for $^{214}$Po; 
$9.4\times 10^4  / 43 $ ($1.1\times 10^5 / 86$ )
with (without) energy cuts for the first $^{212}$Po source;
$1.4\times 10^5  / 43 $ ($1.7\times 10^5 / 173$)
with (without) energy cuts for the second $^{212}$Po source.
 
The s$^{214}$Po mean-life obtained this way is ($236.00 \pm 0.43)$~$\mu$s
with energy cuts, while the measured mean--lives for the 
two samples of  $^{212}$Po are ($426.9 \pm 1.4$)~ns and ($424.5 \pm 1.2$)~ns
with energy cuts, and ($427.2 \pm 1.3$)~ns and ($424.6 \pm 1.1$)~ns without
energy cuts.

\subsection{Analytical fit}
\label{subsec:fit}
In this approach, no energy cuts have been applied to the data, which 
are fit with an analytical model, including an exponential decaying background:
\begin{equation}
\label{eq:exponSignalwBack}
  n_0 e^{-\lambda t} +   b_0 e^{-b_1 t}.
\end{equation}
The method was applied only to the  $^{214}$Po data in the energy window
[0.005--7.605]~ms, since in this case there are both signal dominated and 
background  dominated bins  (see Fig.~\ref{fig:decayIstogram214}).

Total-number-of-events/estimated-background-events in the selected time window
are about $3.9\times 10^5 / 4.4\times 10^4$. No deviation from constant
background is detected.

The so-obtained $^{214}$Po mean-life is ($236.26 \pm 0.47)$~$\mu$s.

This fit and the ones in the previous subsection~\ref{subsec:penalty} are 
performed minimizing the $\chi^2$ using the binned likelihood method.  
Results are stable when changing the number of bins, and the $\chi^2$'s are
compatible with statistical fluctuation around the value expected for the 
appropriate number of degrees of freedom.

\subsection{Unbinned mean value}

For the ideal case of exponentially distributed data with no background and an infinite time window, the average time 
$\sum_{i=1}^{n} t_i/n $, where $t_i$ is the time of the i$^{th}$ event and $n$ the number of events,  is the best estimate of 
the lifetime $\tau$ with the smallest variance $\sigma^2 = \tau^2/n$~\cite{Peierls:1935,Jaffrey:1970a,Jaffrey:1970b}.

\begin{figure}
 \centering
 \includegraphics[width=0.9\linewidth]{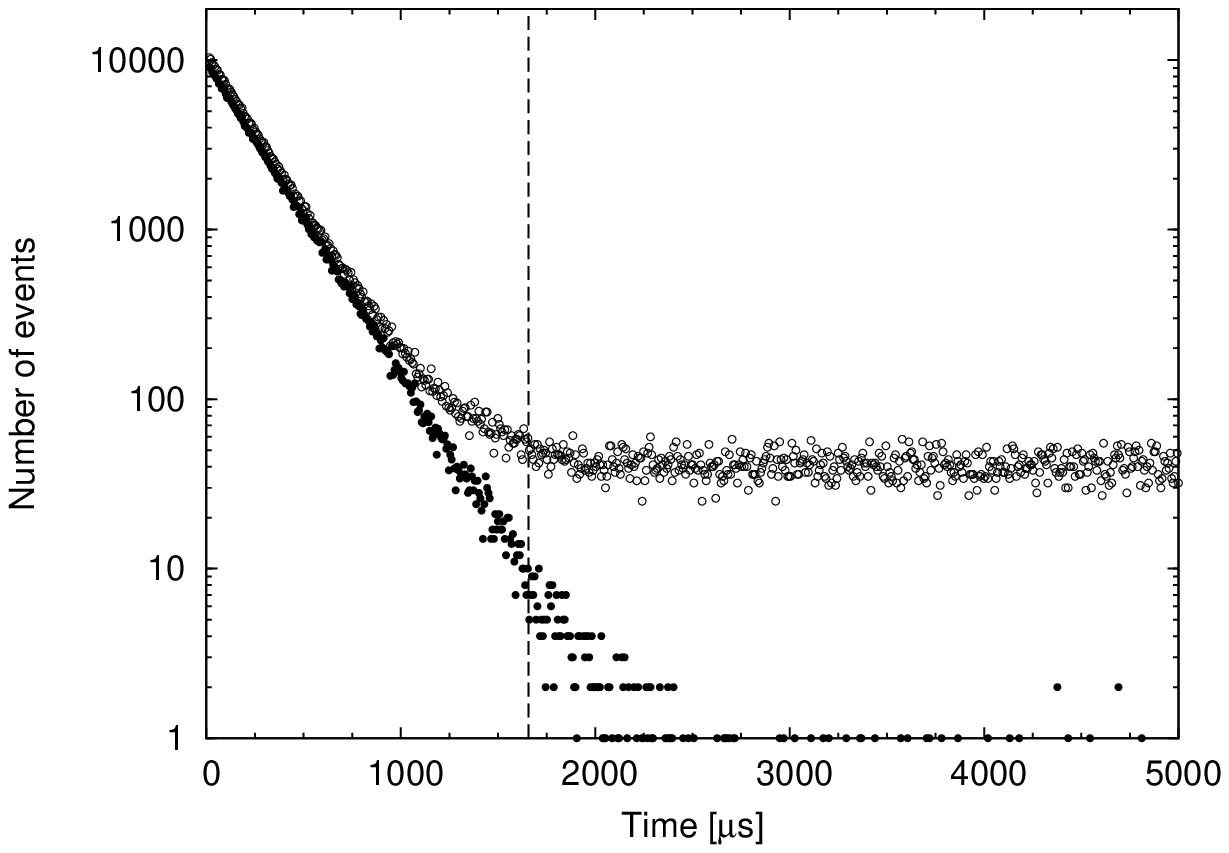}
 \caption{Data from the $^{214}$Po source with 1000 bins between 5 and 7605 $\mu$s 
   are shown up to 5000~$\mu$s. Number of TCE's  as a function of the time difference $t$ between 
 the first and the second decay. Filled (open) dots show  data with (without) 
 the energy cuts reported in the text. 
Vertical lines delimit a seven-lifetime window.
 \label{fig:decayIstogram214}}
\end{figure}

\begin{figure}
 \centering
 \includegraphics[width=0.9\linewidth]{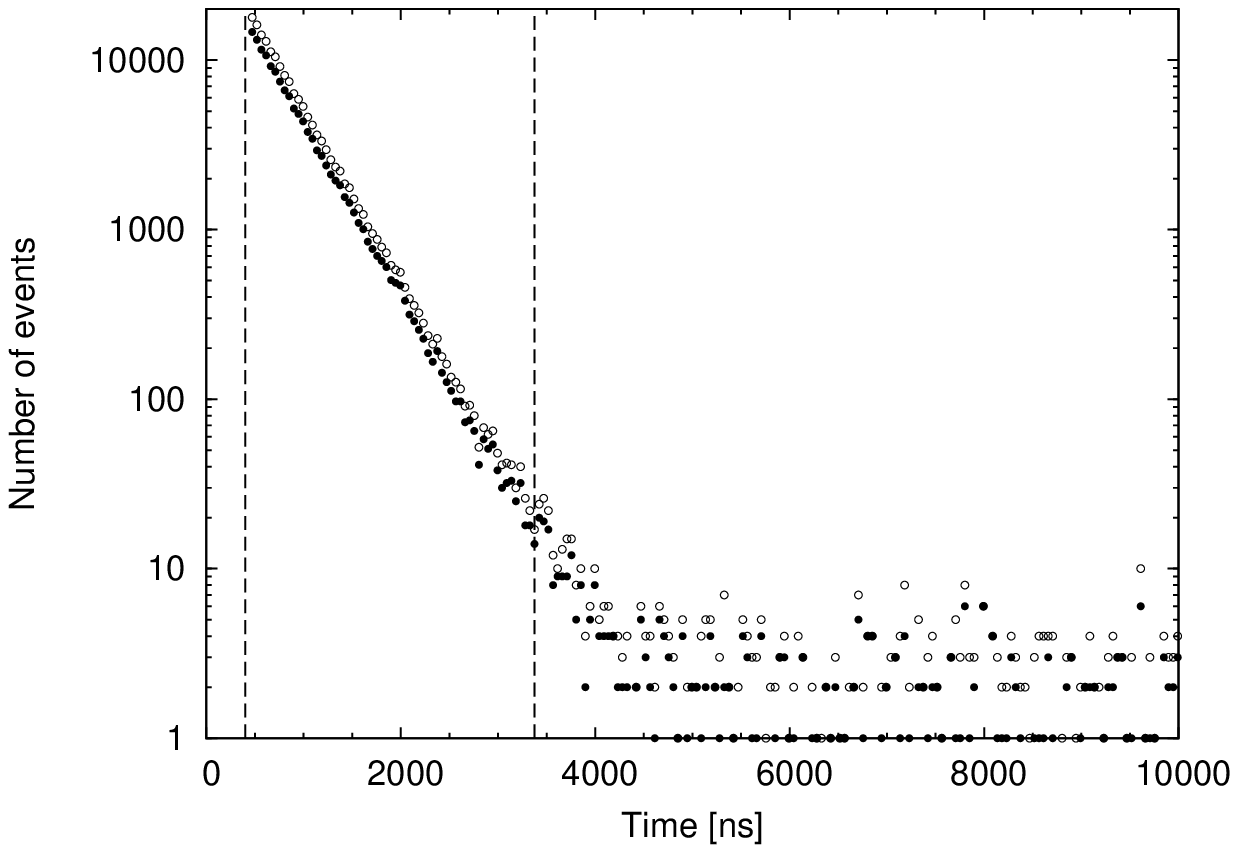}
 \caption{Data from the first $^{212}$Po source with 1000 bins between 400 and 48000~ns are
  shown up to 10000~ns.Number of TCE's  as a function of the time difference $t$ between 
 the first and the second decay. Filled (open) dots show  data with (without) 
 the energy cuts reported in the text. 
 Vertical lines delimit a seven-lifetime window.
 \label{fig:decayIstogram212}}
\end{figure}

Since our data are close to this ideal case, thanks to the favorable signal--to--background ratio, as shown in Figures~\ref{fig:decayIstogram214} and \ref{fig:decayIstogram212}, we can apply the \emph{unbinned mean value} approach with two small corrections: one for the finite width of the time window and the other for a constant background contribution. 

The correction for the finite time window can be calculated analytically: defining $s$ the average decay time in the data window of length $T$,  the relation between the infinite window lifetime $\tau$ and $s$ is: 
\begin{equation} 
\label{eq4}
s= \frac{1}{n}\sum_{i=1}^{n} t_i = \tau - \frac{T}{e^{T/\tau} -1}  \quad .
\end{equation}
A $T\approx  7\tau$ window, used for both polonium measurements, requires a correction of about +0.6~\%.

In order to check the stability of this approach, we split the data in 5 independent equal--size subsets and apply Eq.~\ref{eq4} to each of them. The estimated variance of the mean is $ \sigma^2=\sum_{i=1}^5 (\tau_i -\overline{\tau})^2 /(5(5-1)) $, where $\tau_i$ ($i=1,\ldots, 5$) are the 5 results and $\overline{\tau}$ is their average.

This approach has been implement first without energy cuts, and then applying the same energy cuts already reported 
in Subsection~\ref{subsec:penalty} for the $\chi^2$ with penalty approach. The largest correction for the background is found for the $^{214}$Po source without energy cuts, where the signal--to--background ratio (S/B) is $\approx 36$. The subtraction of a constant background with a number of events about 2.7\% of the signal implies a correction to the lifetime of about -7\%, which is then equal to ($236.85 \pm 0.42$)~$\mu$s. This correction is reduced to about $10^{-4}$ with the energy cuts (S/B $\approx 1.82 \times 10^4$), resulting in a $^{214}$Po mean--life equal to ($236.00 \pm 0.42$)~$\mu$s.

Background corrections for the $^{212}$Po sources are of the order of $10^{-4}$ even without energy cuts. The results for the first sample, 
with (S/B $\approx 2.2 \times 10^3$) and without (S/B $\approx  1.3 \times 10^3$)  the energy cuts, are ($426.9 \pm 1.4$)~ns and ($427.2 \pm 1.5$)~ns, respectively. The same analysis on the second   $^{212}$Po sample leads to a slightly lower values: ($424.6 \pm 1.1$)~ns and ($424.6 \pm 1.1$)~ns. In this case, the S/B ratios are  $\approx 3.3 \times 10^3$ with and $\approx 1.0 \times 10^3$ without the energy cuts.  

Time windows for signal and background and corresponding numbers of events are the same reported in  
Subsection~\ref{subsec:penalty} where the $\chi^2$ with penalty approach is discussed.


\section{Systematic Errors}

The  models, described in the previous section, do not include several detector effects. 
In particular, the PMT jitter ($\lessapprox $1.5~ns, which is  the quality threshold for selecting PMT's in CTF~\cite{Ianni:2005}),  
photon scattering, absorption and remission, and time binning on samples. Further, the maximum difference ($\approx 40$ cm) in the
 photon paths to reach the PMTs induces  a spread in the time distribution of about 0.8~ns. 
Another smearing is due to the TDC  (LeCroy CAMAC 4208, 8 Channel Wide Range Real--Time), with a time resolution of 1~ns, 
which implies a further $\sigma$ = 0.29~ns. 
A detailed Monte Carlo package  has been developed in order to estimate the translation of these smearing effects into 
an  overall systematic error.

In order to calibrate the Master Clock oscillator of the TDC and to check its stability 
we performed a set of measurements using the same ``GPS
disciplined high precision counter system'' utilized for high precision measurements of the neutrino 
speed~\cite{AlvarezSanchez:2012wg}.
We have found that the actual, measured frequency of the Master Clock Oscillator deviates by 
$-47.5$~ppm from it's nominal value at ambient temperature of 23 $^\circ$C.
The short-term oscillator stability during a two-hour run after
stabilization was better than $\pm 0.08$~ppm.
The long term drift of the central value of the oscillator frequency
was less than $\pm 1.5$~ppm during a 24-hour run.
We have found that the temperature dependency of the oscillator frequency
was $0.05$ ppm per degree.
Furthermore, we have observed a  1-$\sigma$ RMS short-term jitter
in the master oscillator of 15 picoseconds.
Taking into account the observed good stability in these measurements,
the standard factory values, usually declared for precision AT cut 
non-oven-stabilized oscillators, crystal aging and temperature changes, 
our sistematic error due to the TDC is within $\pm 20$~ppm. 
This ``worst case'' scenario implies  absolute errors of 
$\pm 0.005\, \mu$s for $^{214}$Po 
and  $\pm 0.009$~ns for $^{212}$Po measurement.

Another subtle effect is due to the difference in the scintillation-photon time distributions  after a $\beta$ or $\alpha$ decay.

The simulation takes into account all the mentioned effects and yields an overall systematic shift of about ($+0.4\pm0.1$)~ns for the
unbinned mean value approach. Therefore, $0.4$~ns are subtracted from our final results for $^{212}$Po (see Table~\ref{tab:res}).

We further studied the result stability by varying the lower threshold of the time acquisition window, from 0.2 to 80~$\mu$s and 
from 0.35 to 0.65 $\mu$s for the $^{214}$Po and $^{212}$Po cases, respectively; in addition, we varied also the upper limit for 
the $^{212}$Po time window from 3 to 3.5 $\mu$s. We also
varied the energy cuts: the lower (higher) charge threshold of the first event in the range 200-300 p.e.
(750-850 pe.e.) for $^{214}$Po, and with and without energy cuts on the first event for $^{212}$Po; the lower (higher) 
charge threshold of the second 
event in the range 130-150 p.e. (180-280 pe.e.) for $^{214}$Po, while we did not apply any cut for $^{212}$Po. 
We adopted a conservative definition of the systematic error as half of the spread of the measured  
lifetimes obtained by applying such different data selection criteria.  The overall estimated systematic 
errors are 0.15~$\mu$s and 1.2~ns for $^{214}$Po and  $^{212}$Po, respectively.

\section{Results and comparison with previous measurements}

The different approaches, used in this analysis, provide compatible results within 1.5~$\sigma$, as shown 
in Table \ref{tab:res}. 
The best results are obtained  by means of the \textit{unbinned mean value} method with energy cuts, 
which provides the smallest statistical uncertainties; $0.4$~ns systematic has been subtracted from
the the $^{212}$Po values.
Since the measurements with the two $^{212}$Po sources are statistically independent, 
we combined the results with the weighted average. 
The final $^{214}$Po and $^{212}$Po mean-life results are :

$$
 \tau(^{214}Po) = 236.00 \pm 0.42 \mathrm{ (stat) }   \pm 0.15 \mathrm{ (syst) } \,  \mu\mathrm{s}
$$
$$
 \tau(^{212}Po) = 425.1 \pm 0.9 \mathrm{ (stat) }   \pm 1.2 \mathrm{ (syst) } \, \mathrm{ns}.
$$ 

We report  the comparisons of our results with the best measurements found in literature  in Tables~\ref{tab:bi214} 
and \ref{tab:bi212}, where statistical and systematic errors are combined.

The $^{214}$Po lifetime measured by von Dardel~\cite{vonDardel:1950} is as accurate as our measurement, although  based on 
larger statistics, corresponding to $\approx  3\times10^6$ events, and with the same acquisition window time length (7 lifetimes).   
The net advantage of the  CTF  measurement is the favorable signal--to--background ratio, higher  by more than 3 order of magnitude.

All available experimental data on $^{214}$Po are reviewed and combined in the Christe et al.  \cite{Christe:2007}, and  Wu et al. \cite{Wu:2009} 
compilations. The first   one combines  all the existing  measurements, finding $\tau = 234.1  \pm  1.7  $~$\mu$s. The second takes into account only 
the three most precise measurements \cite{vonDardel:1950,Dobrowolski:1961,Erlik:1971}, obtaining $\tau = 237.0  \pm 2.9\, \mu\mathrm{s}$.
Note that both compilations rely on techniques for evaluation of data that include some form of Limitation of Relative Statistical Weight
preventing a single measurement to dominate the result: if necessary, the smallest
error is increased  so that the corresponding weight is at most 0.5. 
This is the reason the resulting average and error are not dominated by the one of von Dardel~\cite{vonDardel:1951}, 
which would have by far the smallest error. After including the present measurement 
in these compilations, the average and the error should be determined almost only by the two measurements with smallest errors, 
ours and von Dardel's, with comparable
weights, reducing the final uncertainty on $^{214}$Po lifetime by a factor $\approx 6$ (the adopted number and error depending on the evaluation
technique). A recent experimental test of the time stability of $^{214}$Po half-life~\cite{Alexeyev:2011rw} 
has not yet studied the systematic error of the absolute value. 
If one uses the discrepancy of their results from two difference sets of data as an estimate
of the error, the resulting mean life, $\tau = 235.9  \pm  1.1 \,  \mu$s, is compatible with ours.

Similarly, in the case of the $^{212}$Po, seven measurements have be found in literature \cite{Bunyan:1949,Flack:1962,Astner:1963,McBeth:1972,Sanyal:1975,Bohn:1981}, as shown in Table~\ref{tab:bi212}. Our result  agrees with the most accurate measurement by Sanyal~\cite{Sanyal:1975}, where the statistical 
sample is comparable ($\approx 2\times 10^5$ events), but the acquisition time window is limited to about 4 lifetimes and the signal--to--noise ratio is poorer.  All the other measurements show slightly higher central values, with larger uncertainties, and compatible with the present work at 2 $\sigma$ level.

In the work by Browne et al.  \cite{Browne:2005} the average of five measurements \cite{Bunyan:1949,Astner:1963,McBeth:1972,Sanyal:1975,Bohn:1981} 
is quoted and it is equal to $\tau = 431  \pm 3 $ ns. 
Another work by Nichols et al. \cite{Nichols:2004} takes into account six 
measurements \cite{Bunyan:1949,Flack:1962,Astner:1963,McBeth:1972,Sanyal:1975}, finding  $\tau = 433  \pm 32$ ns.

As in the $^{214}$Po case, evaluation techniques avoid that a single measurement has a weight larger than 0.5. The present result, when 
included in these compilations, should give the more important contribution together with the measurement of Sanyal \cite{Sanyal:1975} to
the adopted lifetime, which will become lower, and to the uncertainty, which will be reduced.

In conclusion,  thanks to extreme radio-purity of the CTF detector and to a long expertise in source preparation and insertion systems, new accurate measurements of the $^{214}$Po and $^{212}$Po lifetimes have been provided. 

\acknowledgments
We acknowledge the generous support of the Laboratori Nazionali del Gran Sasso and we thank the 
funding agencies: INFN (Italy), NSF (USA), BMBF, DFG and MPG (Germany), Rosnauka (Russia), 
the Ministry of Education and Science (Russia), and MNiSW (Poland).
We are grateful for enlightening discussions with and
the valuable comments of E. Bellotti and B. Ricci.

\end{document}